\documentclass[12pt]{article}

\usepackage{amsmath,amsfonts,epsfig,color,latexsym}

\newcommand{\be}{\begin{equation}}
\newcommand{\ee}{\end{equation}}
\newcommand{\bea}{\begin{eqnarray}}
\newcommand{\eea}{\end{eqnarray}}

\renewcommand{\theequation}{\thesection.\arabic{equation}}

\let\newsection=\section
\renewcommand{\section}{\setcounter{equation}{0}\newsection}

\begin{document}

\begin{flushright}
hep-th/0603176\\
BROWN-HET-1466
\end{flushright}
\vskip.5in

\begin{center}

{\LARGE\bf More on the RHIC fireball and dual black holes }
\vskip 1in
\centerline{\Large Horatiu Nastase}
\vskip .5in

\end{center}
\centerline{\large Brown University}
\centerline{\large Providence, RI, 02912, USA}

\vskip 1in

\begin{abstract}

{\large We revisit the issue of the RHIC ``fireball'' as a dual black 
hole, and explain some of the details. We discuss the nature of the 
(black hole) information paradox as a purely field theory (gauge theory)
phenomenon and how the paradox can be formulated in exactly the same way
for the RHIC fireball and a black hole. We stress the differences between 
the black holes produced in the gravity dual and the equilibrium situation 
of the Witten construction for finite temperature AdS-CFT.
We analyze the thermodynamics of the fireball, give more arguments why 
$T_{fireball}\propto m_{\pi}$, including an effective field theory one,
 and explain what entropy=area/4
means experimentally for the fireball.  
}

\end{abstract}

\newpage

\section{ Introduction}

AdS-CFT \cite{malda} relates string theories in gravitational backgrounds 
with gauge theories in flat spacetime. In the simplest case, $AdS_5\times 
S_5$ string theory is related to ${\cal N}=4$ SYM in 4 dimensions and 
one has a rather extensive dictionary for the duality, but in 
cases with less symmetry and more realistic features, the gravitational 
background (gravity dual)
is harder to obtain, and the dictionary is less developed. 
However, Polchinski and Strassler \cite{ps} have proposed a duality dictionary
for computing high energy scattering in QCD from scattering in the gravity 
dual  without knowledge of the 
precise gravity dual, by using a simple model with an $AdS_5\times X_5$ 
space with the $AdS_5$ cut-off in the IR. In a series of papers 
\cite{kn,kn2,nastase2,nastase,nastase3} 
it was shown that one can use a formalism
of scattering of Aichelburg-Sexl (AS) type 
shockwaves in the gravity dual to calculate 
quantitative features of high $s$, fixed $t$ scattering in QCD and find 
various energy regimes. The scattering produces black holes in the 
gravity dual if the collision energy is larger than $M_P$, equivalently if 
the QCD collision energy exceeds $\hat{M}_P=N_c^{1/4}\Lambda_{QCD}$.
It was found that the last energy regime corresponds 
to the saturation of the Froissart unitarity bound in QCD \cite{kn2}, and 
in \cite{nastase} it was proposed that the RHIC fireball corresponds to 
scattering in the Froissart energy regime and that the RHIC fireball 
is dual to a black hole living on the 4d ``IR brane'', the flat cut-off of 
AdS space (for different ideas on how to relate black holes to RHIC collisions,
specifically in the context of ${\cal N}=4 SYM$, see \cite{sz,shuz,ssz,jp}). 
In this paper we will clarify some of the issues relating to 
the duality of the fireball to 
black holes living on the IR brane, in particular the properties of the 
produced dual black holes.

The formalism of scattering two gravitational AS shockwaves to create black 
holes was proposed in \cite{eg} and extended in \cite{kn3} to nonzero 
impact parameter $b$ in any dimension and to curved space, making it 
useful for calculations of scattering in gravity duals (see also 
\cite{nastase4}). At fixed $t$ and 
high enough $s$ ($s>M_P$), classical scattering of gravitational shockwaves
dominates the quantum scattering and one can calculate that there is a 
trapped surface forming, thus by general relativity theorems one knows there 
is a horizon larger than the trapped surface forming in the future of the 
collision, i.e. a black hole. At high enough $s$ in the gravity dual
of QCD (corresponding to 
the saturation of the Froissart bound in QCD), the scattering takes place 
effectively on the IR brane, thus one creates a black hole on the IR brane.
The trapped surface created in the AS shockwave collision at zero impact
parameter corresponds (in a 
particular set of coordinates) to two flat disks of radius \cite{kn}
\be
r_{H,ap}\simeq \frac{1}{M_1}ln [R_sM_1\frac{3C_1}{M_1R}\sqrt{\frac{\pi}{M_1R}}
]\simeq \frac{1}{M_1}ln [R_s M_1]
\ee
where $R_s=2G_4\sqrt{s}$ and $M_1= j_{1,1}/R\simeq 3.83/R$ is the mass of the 
first KK excitation if one dimensionally reduces onto the IR brane. 
We can deduce that the area of the horizon of the 
formed black hole satisfies
\be
A_{horizon}
\geq 2\pi r_{H,ap}^2\simeq \frac{2\pi}{M_1^2}ln^2 [R_sM_1]
\ee
Doing the calculation at nonzero impact parameter, one finds that the 
maximum impact parameter that produces a trapped surface is 
\be
b_{max}(s)\simeq \frac{\sqrt{2}}{M_1} ln [R_sM_1(\frac{3\sqrt{\pi}}{\sqrt{2}
j_{1,1}^{3/2}})]
\ee
giving a cross section for black hole creation of 
\be
\sigma= \pi b_{max}^2 \simeq  \frac{2\pi}{M_1^2} ln^2 
[R_sM_1(\frac{3\sqrt{\pi}}{\sqrt{2} j_{1,1}^{3/2}})]\simeq 
 \frac{2\pi}{M_1^2} ln^2 [R_sM_1]
\ee
so that to leading order (at large $s$) the cross section for black hole 
creation approximates the area of the trapped surface formed at zero impact 
parameter. 

At this point we should mention that the purely gravitational scattering 
analyzed here is actually dual to pure Yang-Mills (no light quarks), so 
the lightest excitations of the gauge theory are glueballs of mass $m$, dual 
to $M_1$. In reality, the lightest QCD 
excitation is the pion, made up of light 
quarks, and the pion field is dual to the position of the IR brane in the 
fifth dimension (brane bending)
\cite{gid,kn,kn2}, but it behaves in a manner analogous to gravity.
However, it is easier to calculate the scattering in the purely 
gravitational case, so in most of the rest of the paper we will talk about 
pure gravity. 

The dimensionally reduced gravity onto the IR brane is massive with mass 
$M_1$, meaning that perturbative 4d gravity (at large distances $r$ from the 
source) gives for a point source of mass $M=\sqrt{s}$
\be
h_{00}=1+g_{00}\sim \frac{2G_4\sqrt{s}e^{-M_1r}}{r}=\frac{R_se^{-M_1r}}{r}
\label{4dpert}
\ee
that means that the perturbative position of the horizon would be where 
$h_{00}\sim 1$, giving 
\be
r_{H,pert}\simeq \frac{1}{M_1}ln [R_sM_1]\Rightarrow \sigma_{pert,geom}=
\pi r^2_{H,pert}\simeq \frac{\pi }{M_1^2}ln[R_sM_1]
\ee
that is, the geometric projected area of the perturbative horizon has
the same leading behaviour (up to a multiplicative constant) as the 
cross section for black hole formation and the area of the trapped surface
formed at $b=0$. One would be inclined to say that the area of the actual 
horizon created in the collision would behave similarly, however we argued 
in \cite{nastase} and will further explain in the following 
 that this is not the case, and 
thus the scattering cross section behaves in this case differently than the
area of the actual horizon (but the same as the area of the
perturbative horizon). In particular, we will later on 
argue for the fact that the 
area of the actual black hole horizon is proportional to its mass. 

\section{ Thermal decay in QFT and the information paradox}

Let us examine in detail some qualitative features of 
the fireball - dual black hole map. It has been clear for a long time 
(since the beginings of AdS-CFT) that if the duality holds in its most 
general form, a black hole should have some dual description in terms of 
gauge theory objects. The original proposal of Witten \cite{witten} 
identified a black hole in AdS space as corresponding to putting the 
dual field theory at finite temperature. But the process of creation and 
evaporation of a black hole (non-equilibrium thermodynamics)
should also be described, and people came with 
various proposals on how to do this (see e.g., \cite{vijay}). 

The proposal in \cite{nastase}
however states the problem in a very clear way: The Polchinski-Strassler 
formalism relates high energy scattering in the gravity dual to high 
energy scattering in the gauge theory, and scattering of {\em any kind of 
light particles} at $s>M_P$ in the 
gravity dual creates black holes, that will then decay through Hawking 
radiation. In the gauge theory, that corresponds to the fact that scattering
of {\em any kind of hadrons or glueballs}, thus in QCD 
not only nuclei as at RHIC but also nucleons (p and n) should produce an 
object that decays thermally (what we called a ``nonlinear pion field 
soliton'' in \cite{nastase2,nastase} and \cite{amw} calls a ``plasmaball''). 

The argument of 't Hooft \cite{thooft,kn3} for $s$ less than, but close to 
$M_P$ (massive interactions are irrelevant, since they are infinitely 
time delayed, and massless interactions are encoded in the shockwave 
scattering), as well as explicit calculation of string corrections to 
scattering \cite{kn3}, guarantee
 that AS shockwave scattering gives the correct leading result. The 
calculation of $g_s$ and $\alpha '$ string corrections to the fixed $t$, 
large $s$ AS shockwave scattering in the gravity dual
 \cite{kn3,kn2,nastase2} (they are exponentially small in $\sqrt{s}$)
guarantees that $g^2_{YM}$ 
and $1/N$ corrections to the scattering are small in 
gauge theory, thus one can apply the formalism to QCD. Nevertheless, even 
if one has doubts over whether black hole production actually dominates the 
QCD total cross section at fixed $t$, large $s$ (for small $\lambda=g^2_{YM}
N$) \cite{amw,tan}, the fact is that one {\em has to} produce dual objects
in gauge theory that decay thermally. Let us see why this is so.

The existence of such thermal objects that can be created in the 
quantum (unitary) collisions of hadrons and then decay with aparent 
loss of unitarity underscores the fact that the information paradox is not 
a gravitational issue, rather it is a quantum field theory phenomenon. 
All we need for that is AdS-CFT a la Polchinski-Strassler 
and the statement that an aparent loss of unitarity occurs in the formation 
and decay of a black hole to deduce that apparent loss of unitarity occurs 
in gauge theory in a certain process of formation and decay of a metastable 
state, {\em independent of whether this process dominates the scattering}. 
In other words, the aparent information loss due to Haking radiation 
is just a statement in the dual gauge theory
about the lack of a formalism to deal with the formation 
and decay of a thermal object. 

In quantum field theory one usually deals with finite temperature by imposing 
it by fiat (i.e., there is a constant finite temperature throughout space and 
time- it cannot vary in space, nor is it created in time). 
Yet the RHIC experiments are an example of a case where we 
produce a thermal quantum field theory system (the ``fireball'', localized 
in space and time)
via the collision and decay of an essentially quantum system. The problem 
is obscured by the fact that the collision involves nuclei rather than 
nucleons, specifically Au on Au (each with about 200 nucleons) at about 
200 GeV/nucleon. Thus one might believe that due to the large number of 
nucleons the process of thermalization is essentially a classical one
(colliding two groups of classical balls, at rest within each group, will 
certainly generically create a thermal system after the collision). 
Classical thermalization is no mistery, and if we try to describe this 
system in quantum mechanics, the apparent loss of unitarity in going 
from a pure state to a mixed state is not more than the usual decoherence
phenomenon (also apparently non-unitary)
in going from a quantum state to a classical state. 
But the same could be said about the creation and decay of a black hole.

Of course, for a black hole the point is that the radiation coming out 
is completely thermal, and completely independent of the objects coming in, 
i.e. apparently there is no information being kept. In other words, if we 
collide ``classical balls'' of type $A_j$ to create a black hole, the 
black hole will radiate all possible $\{A_i \}_{i=1,n}$'s 
in a thermal distribution, 
independent of the initial $A_j$. But exactly the same is happening in the 
process at RHIC! The decay products are all thermally distributed, with the 
lightest particles, the pions, contributing predominantly, independently of 
what is it that we collide. This point is slightly obscured by the fact that 
one collides always the same nuclei in the experiment, but consider that the 
collision involves neutrons and protons, and the decay products are mostly 
the lightest particles of the theory, the pions! 

The nuclei collision creates 
the hot fireball, that then decays through hadronization into mostly 
lightest particles = pions, 
exactly as the collision of massive objects would form a gravitational 
black hole, that then radiates thermally mostly massless particles.
The thermalization (and aparent loss of unitarity) really occurs during 
the formation and decay of the hot fireball, when protons and neutrons 
become a plasma, and then rearrange into pions, a fact underscored by the 
observation that the number of pions emitted $N_{\pi}$
is also independent on the number of initial particles ${\cal A}$, 
but rather depends on the total energy of the collision $\sqrt{s}_{tot}$.
This is again obscured in the experiment, as the number of inital nucleons 
participating in the reaction, ${\cal A}$, is proportional to the energy 
of the collision, $\sqrt{s}_{tot}$, but that is easily correctable if one 
varies the energy per nucleon independently in the experiment.
Yet it is intuitively clear this is so, as the number of emitted pions 
is in the tens of thousands, and the number of nucleons of the order of a 
hundred. 

In conclusion, the formation and decay of the RHIC fireball is as misterious 
as the black hole information paradox, and one lacks a proper (purely)
QCD description of how the fireball forms and decays same as one lacks a 
quantum description of the black hole. Specifically, {\em the lack of 
a quantum field theory 
formalism describing non-equlibrium thermodynamics} (formation and decay 
of a thermal object inside a T=0 vacuum) is mapped to the black hole 
information paradox. But like there is 
the quasi-classical description of a classical black hole decaying by 
Hawking radiation, we have put forward in \cite{nastase3} a quasiclassical 
description in effective field theory for the fireball. There is a more 
detailed analysis, but one can model the physics with a simple scalar 
DBI action, $\int \sqrt{1+(\partial_{\mu} \phi)^2/\Lambda^2}$. 
The action for the scalar,
standing for the pion field, has a solution (``catenoid'')
with an apparent singularity at $r=r_0$, which in $\Lambda$ units is
($\Lambda$ for QCD is of the order of $\Lambda_{QCD}$, or $\hat{M}_P$)
\be
\phi'=\frac{\bar{C}}{\sqrt{r^4-\bar{C}^2}}=
\frac{r_0^2}{\sqrt{r^4-r_0^4}}
\label{dbiscalar}
\ee
and the apparent singularity (where $\phi '$ is infinite, but $\phi$ is finite)
has the same properties as a horizon: It radiates particles thermally, exacly 
as the hydrodynamic ``dumb holes'' of Unruh have horizons for the speed of 
sound, where $v=c$, that radiate thermally (see fig.\ref{scalar}).
These semiclassical ``catenoids'' unfortunately 
have infinite temperature \cite{nastase3}, 
but if the pion has a mass $m$, the temperature 
is proportional with $m$ (with an infinite factor multiplying it). Making the 
temperature finite by some regularization (or by the fact that one has to 
take a purely quantum treatment, not semiclassical as above)
would at the same time make the travel time for {\em information}
 to reach the horizon (or come from it) infinite, same as for a black hole. 
The possible 
continuation of the solution inside the horizon has been analyzed in 
\cite{nastase3}, and the only consistent possibility seems the metastable 
situation depicted in fig.\ref{scalar}, which is easy to understand: in a 
collision we create a field that perturbatively (at large distances) looks 
like (\ref{dbiscalar}), thus has a horizon, and to connect with the trivial
profile ($\phi$=constant) before the collision, the only possibility inside 
the horizon is the one depicted.

\section{ Solving apparent contradictions with QCD picture}

There are a number of questions that were posed to me since the 
appearence of \cite{nastase} about the picture of the fireball as a dual 
black hole that I would like to clarify, since they weren't very clear 
in \cite{nastase}.

\begin{figure}

\begin{center}

\includegraphics{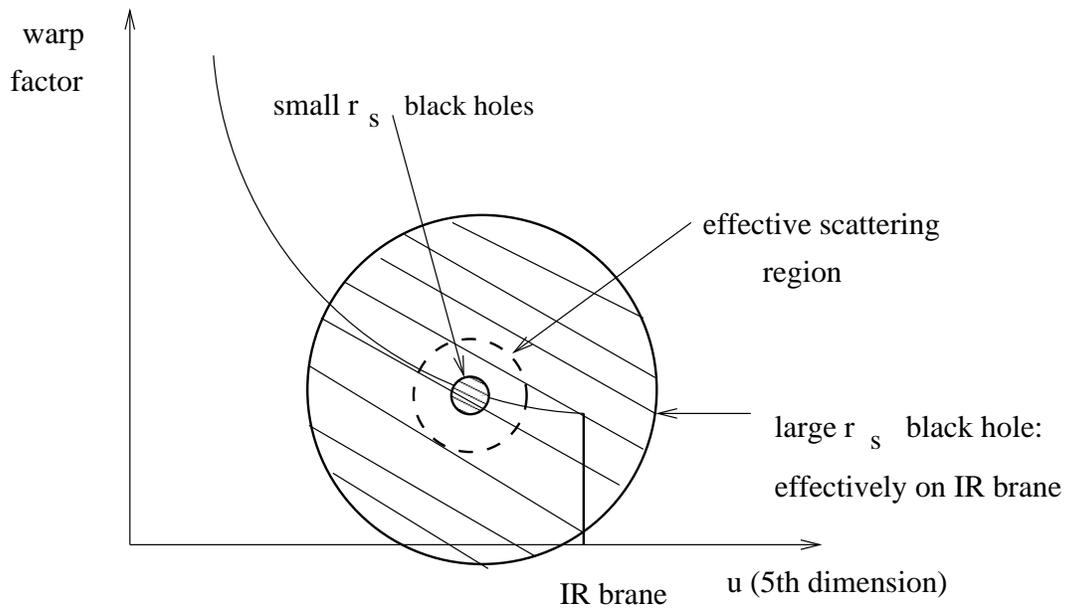}
\end{center}
\caption{Scattering in the cut-off AdS gravity dual. Small black holes 
are effectively produced in a scattering region and don't feel the IR brane.
Asymptotically large black holes live effectively on the IR brane and have 
negligible quantum fluctuations in the 5th dimension. }
\label{region}
\end{figure}

1. A black hole is ``black'', i.e. everything that goes in gets converted 
to thermal radiation, and cannot get out as is, 
but experimentally the fireball is not perfectly absorbtive. The ``jet 
quenching'', the diminishing of jets (hard scattering events) due to absorbtion
by the fireball, is not complete. This is easily understood in our picture, 
since the dual black hole lives only effectively on the IR brane. Scattering 
in QCD corresponds to scattering in the gravity dual, with the amplitude 
being integrated over the extra dimensions, $
{\cal A}_{gauge}\sim \int dr d^5\Omega \sqrt{\hat{g}} {\cal A}_{string}\prod_i
\psi_i$, and it is dominated by  a region close to the IR brane, but otherwise
can be anywhere. Thus a particle can miss the black hole because it is at 
a different point in the extra 
dimensions than the test particle,
due to quantum fluctuations. Only in the extreme large mass limit, when 
the black hole is effectively on the IR brane and classical, and very large,
will the absorbtion be total, but then the same will happen in QCD (see 
fig.\ref{region}). However, at finite mass it is very hard to calculate 
the exact jet quenching without knowing the exact QCD gravity dual.

2. The collision at RHIC involves two ultrarelativistic nuclei, Lorentz 
contracted to look like pancakes (shockwaves in the limit), creating an object 
that is initially very flat, then expanding to a ellypsoidal shape while 
starting to decay, and then decaying via hadronization, and as a result there 
is a complicated angular distribution of the decay products. But the black 
hole considered here is spherical in 4d. Of course, the collision of two 
gravitational shockwaves will also create a shockwave (``pancake'')
object that will 
start expanding towards a spherical shape. The problem is that the metric 
in the future of the shockwave collision is unknown, and one can only 
calculate the creation of a trapped surface. So we took as a model the 
creation of a static spherically symmetric dual black hole, but in 
principle one should calculate (at least numerically) the time 
evolution of the shockwave collision. 

3. The temperature of the black hole was conjectured to be \cite{nastase}
 $T=4aM_1/\pi$, and correspondingly the temperature of the fireball is 
$T= 4a<m_{\pi}>/\pi$, independent on the decay products, i.e. the temperatures
of the pions, kaons, etc. will be the same, and depend only on the  pion mass, 
as the lightest excitation of the theory (the quark content of the emitted 
particles should be irrelevant, only their mass should matter). 
But experimentally, there might be a 
better fit if one assumes different temperatures for the different particles.
But again, this is a dynamical process of an expanding pancake-like object, 
and while a static black hole would give the same temperature for all the 
decay products, an evolving black hole could give different temperatures for
various products, all that matters is that the asymptotically the
average temperature is independent on the details of the experiment, and 
only depends on the pion mass.

4. The calculated temperature of the fireball is only the temperature of 
the decay products, but the result (both our theoretical and the experimental
value) is about 175 MeV, which happens to be the same (within errors) as the
temperature of the phase transition, calculated on the lattice. But why 
would the dual black hole temperature correspond to a phase transition 
temperature? This was explained in \cite{amw}, and we will comment more on it
later on. The point is that the black hole looks also like an object dominated
by a different phase (encompassing most of its interior), thus it 
thermodynamically corresponds to a transition between a graviton gas and 
the new phase, hence its asymptotic temperature is the temperature of the 
phase transition.

5. Apparently, when you take the pion mass to zero, the temperature of the 
fireball, which we saw is related to the phase transition temperature,
 should go to zero. How is this possible?  The point is that all the 
calculations fail in the $m_{\pi}\rightarrow 0$ limit, corresponding to 
$M_1\rightarrow 0$. First, this is the flat space limit ($R^{-1}\rightarrow 0$)
of the gravity dual, 
so the dual becomes just flat space and the IR brane loses its meaning. 
Instead of black holes on the IR brane, with a different phase inside them, 
we would have black holes in flat space, with no different phase. In QCD, that
will manifest itself in the fact that the Froissart bound will dissappear, as 
the coefficient in  $\sigma_{tot}(s)\leq A\;
ln^2s$ satisfies $A\leq \pi/m_{\pi}
^2\rightarrow \infty$, so we will never reach the Froissart saturation
regime. However,
that only says that the asymptotic regime, when the fireball is dominated 
by the new phase inside, is becoming harder and harder to attain (higher
energy needed), but if the asymptotic 
fireball measures the temperature of the phase transition (and $T_{fireball}
\propto m_{\pi}$), it would imply that the phase transition temperature 
will also go to zero. But we have forgotten that we need to keep something 
fixed in this limit. We could keep $M_P$ fixed, 
corresponding to $\hat{M}_P=N_c^{1/4}
\Lambda_{QCD}$ in gauge theory, but for $N_c=3$ that is the same as keeping 
$\Lambda_{QCD}$ constant, so it doesn't help. Therefore for this limit, we 
cannot consider just Yang-Mills fields as we did until now
(and define the pions as the lightest 
excitations), we have to take Yang-Mills fields and light quarks, composing 
the pions.

\begin{figure}

\begin{center}

\includegraphics{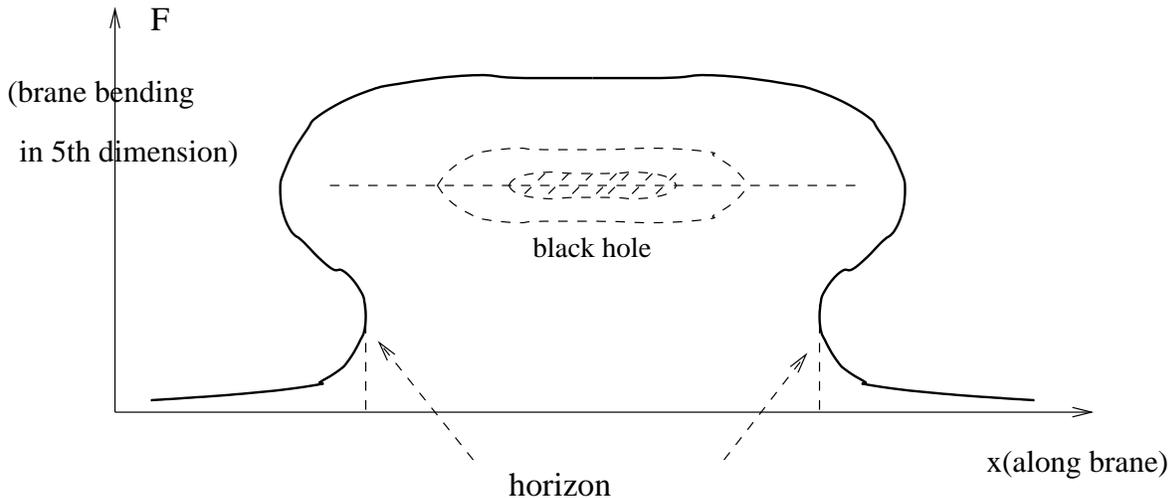}
\end{center}
\caption{Brane bending profile, corresponding to pion field 
profile (case when brane bending mass is lowest). It also has a ``horizon''
(aparent singularity in the field), but inside we have a smooth continuation
to a different asymptotic region. The formed black hole (dotted contour)
will be smaller, thus fit into the new asymptotic region of F
(or $\phi$ as is denoted in the text), and the 
black hole core will correspond to a different phase. }
\label{scalar}
\end{figure}

The picture then is more complicated. Brane bending in the gravity dual
starts dominating before 
the black hole created in the gravity dual can reach the IR brane. The 
asymptotically large fireball -its thermodynamics as well- has to be 
dominated by the light pion field (brane bending in the dual), so 
one would get $T_{fireball}\propto m_{\pi}$. This was argued in the 
simple model in \cite{nastase3} (even though the simple model there
gave an infinite multiplicative factor). However, if the pion mass (brane 
bending mass) $m_{\pi}$ becomes much smaller than the lightest glueball 
mass (KK graviton mass) $M_1$, eventually black hole creation 
in the dual becomes irrelevant and only brane bending will matter (see 
fig.\ref{scalar}). 
In QCD that would mean that 
the fireball will still have $T_{fireball}\propto m_{\pi}$, as we will argue
further in section 5, but that it will 
not have the new state of matter (coming from KK gravitons in the dual, 
governed 
by $M_1$) inside the fireball, thus the phase transition temperature will
decouple from the fireball temperature (``freeze-out'' temperature of the 
decay products), and remain finite as we take $m_{\pi}$,
thus $T_{fireball}$, to zero. 
Brane bending will create the metastable shape depicted in fig.\ref{scalar},
which has the correct asymptotics, with a horizon, and the continuation inside
the horizon is the most plausible, as we already argued. 
Of course, this picture needs to be confirmed 
by calculations in the case of both black holes and brane bending being 
created, but it matches also the rough picture observed on the lattice.
Specifically, on the lattice one finds that as the quark mass goes from 
their physical value to zero, 
the phase transition temperature drops from about 175 MeV to about 150 MeV.
That would mean that the physical value for $m_{\pi}$ is very close to 
the point where the fireball (``freeze-out'') temperature decouples from 
the phase transition temperature. Of course, we have to have also an upper
bound for $m_{\pi}$ for the dependence of the phase transition temperature 
on $m_{\pi}$, since as $m_{\pi}$ becomes close to $M_1$, the physics will 
start being dominated by $M_1$, not $m_{\pi}$. It will be interesting to 
see if lattice calculations can shed more light on this picture. 

6. There is no singularity inside the fireball, so the black hole on the 
IR brane shouldn't have one either, why? In the gauge theory side it is 
clear that the fireball will not have a singularity at its center, so we 
expect that the black hole on the IR brane will also not have a singularity 
at its center. In fact, this was the picture found in \cite{amw}. 
The black hole on the IR brane looks like a ``pancake'' (fig.\ref{IRbrane}), 
with most 
of the horizon being parallel to the brane. If there would be a singularity,
it would have to be extended along the brane as well. The AdS background 
\be
ds^2=R^2[e^{2u}(-dt^2 +d\theta^2 +d\vec{x}^2)+du^2]
\ee
has to be modified in the IR at some finite 
$u_0$, corresponding to the IR brane. In \cite{amw} such an example of 
background (gravity dual) is 
presented (the ``AdS soliton''), together with a black brane with translational
invariance in $\vec{x}, \theta$, that is also a deformation of AdS,
with horizon at finite u and a singularity at $u=-\infty$. The black hole on
the IR brane is a solution that outside the horizon interpolates between 
the AdS soliton and black brane, but since most of the space terminates at 
$u_0$, the interior cannot continue to a $u=-\infty$ singularity as the black 
brane does. We will say more on this in the following.

\section{ Black holes produced in the gravity dual}

We will now try to answer the question of how do the black holes created in 
the cut-off AdS look like, and how does that connect to the black holes on 
the IR brane. 

For a gauge theory in equilibrium at finite temperature T, Witten 
\cite{witten} considered as gravity dual 
the Euclidean AdS black hole, that has a compact coordinate with periodicity
defining the temperature of the system. We however want to consider the 
creation and decay of a finite temperature fireball, corresponding to a 
black hole created in the gravity dual. Initially (at small $\sqrt{s}$), 
the black hole is small 
and feels just flat space, then (at larger $\sqrt{s}$)
starts feeling the curvature of space (fig.\ref{region}). 
We will take a simple model of just a cut-off $AdS_5$ (no extra compact 
5d space) and see at the end if that was justified. 

{\em Flat space}.
For a 4d Schwarzschild black hole, the singularity at $r=2M$ is a coordinate 
singularity. In Lorentz signature, the singularity can be removed by 
analytically continuing inside the horizon, in Kruskal coordinates. Hawking
calculated the temperature of the black hole from the analysis of the 
propagation of normal modes and found that $T=k/(2\pi)=1/(8\pi M)$ ($k$= 
surface gravity). 
The calculation in Euclidean space is simpler:
the metric  outside the horizon is 
positive definite, and we cannot analytically continue inside the horizon.
Instead, in Euclidean space ($\tau =it$), the singularity is a conical 
singularity: $\partial /\partial \tau $ is an axial Killing vector. 
Near $r=2M$, the metric is 
\be
ds^2 |_{r\simeq 2M}\simeq 2M[ dy^2 + y^2 (\frac{d\tau}{4M})^2 +\frac{y^4}
{16 \cdot 2M}d\Omega_2^2]
\ee
and $dr^2 +r^2 d\theta^2$ is nonsingular at $r=0$ only if $\theta$ has 
periodicity $2\pi$ (otherwise there is a conical singularity). Thus now, 
there is no singularity only if $\tau$ is periodic with periodicity $8\pi M$.
The periodicity in Euclidean time means a finite temperature, 
T= 1/ periodicity, i.e. $T=1/(8\pi M)$, equal to the Hawking calculation in 
Minkowski space.

By direct calculation, one finds that  the Scwarzschild black hole has 

-negative specific heat $\partial M/\partial T <0$

-positive free energy $F>0$, where  F/T=I, i.e. the value of action on 
the solution.

Both facts imply that the solution is unstable, and indeed the black hole 
gets radiated away. 

{\em AdS space}. Similarly to the above,
for the black hole in AdS space (see the Appendix), one  
has the same Hawking calculation  in Lorentzian signature, getting 
\be
T= \frac{\kappa }{2\pi }= \frac{ \partial_r V|_{horizon}}{4\pi}
\ee
which gives now (using $V$ from the Appendix)
\be
T= \frac{nr_+^2+(n-2)R^2}{4\pi r_+ R^2}
\ee

In Euclidean signature the calculation is also the same: 
we again write the metric near the singularity as 
\be
ds^2= \frac{1}{A}[dy^2 + y^2 (\frac{Ad\tau}{2})^2 +\frac{Ay^4}{16}d\Omega_{n-1}
^2];\;\;\; V|_{r\simeq r_+}\equiv A (r-r_+)=\frac{A}{4}y^2
\ee
thus to avoid conical singularities $\tau \sim \tau + 4\pi / A$, giving 
the same temperature as above.

{\em Thermodynamics}

\begin{figure}

\begin{center}

\includegraphics{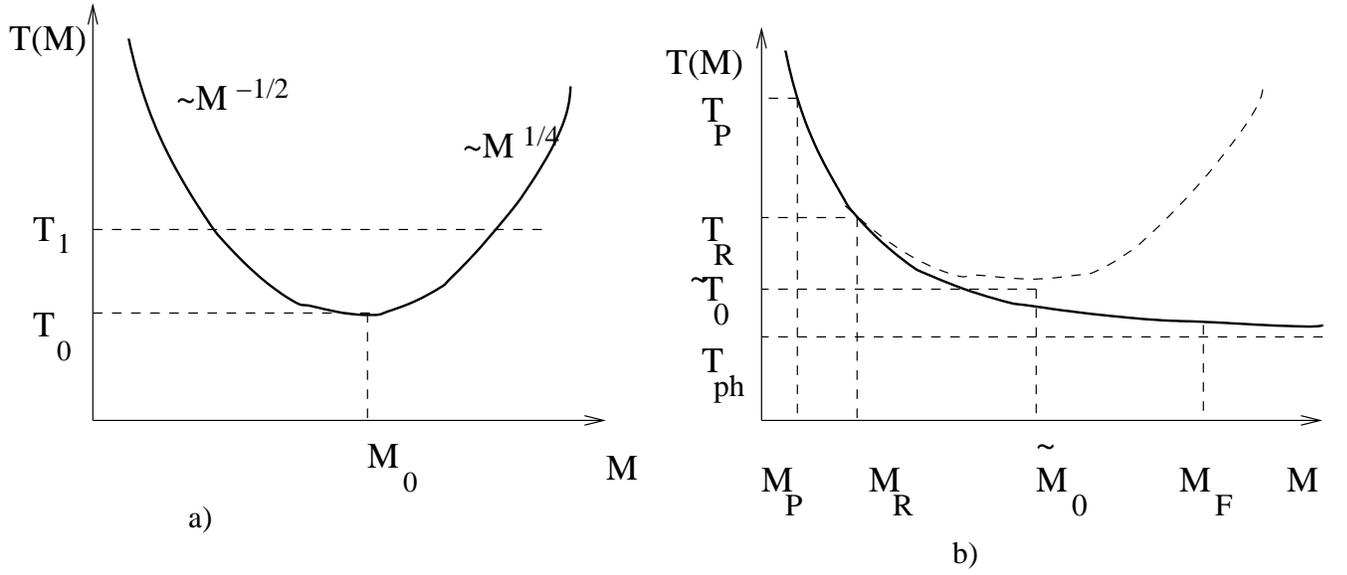}
\end{center}
\caption{a) T(M) for black holes in AdS. The lower mass branch starts off as 
in flat space, goes down to a minimum $T_0$ and then up. Above $T_1$ 
on the large M branch we have 
$F<0$. b) T(M) for scattering in gravity dual. Dotted curve would correspond
to gravity dual without cut-off. Asymptotic value $T_{ph}$ corresponds to 
large black holes on the IR brane. Here $\tilde{T}_0,\tilde{M}_0$ 
correspond to $AdS_5\times X_5$ (compact space taken into account) }
\label{temp}
\end{figure}

From $T(r_+)$ together with the fact that $M$ increases monotonically 
with $r_+$, we get that 
T(M) decreases down to a $T_0$, then increases, as in fig.\ref{temp}a.
  For $AdS_{n+1}$, we have
\be
T_{min}=\frac{\sqrt{n(n-2)}}{2\pi R}=T_0;\;\; r_+=r_0= R\sqrt{\frac{n-2}{n}}
;\;\; M= \frac{r_+^{n-2}}{w_n}(1+\frac{r_+^2}{R^2}); 
\ee
The action of the black hole solution, minus the AdS 
action  is proportional to 
$R^2-r_+^2$, and this is identified as $I-I_0=F/T$ (we have exp(-I) in 
the partition function), thus F=0 when $r_+=R$, hence when $T=T_1= 
(n-1)/2\pi R$.

That means that for a given AdS Euclidean temperature $T>T_0$, 
there are 2 values of M for a black hole to be in equilibrium. The lower
branch is the one that survives in flat (n+1) dimensional
 space. As in flat space, it has 
negative specific heat $\partial M/\partial T<0$ and positive free energy 
$F>0$, thus is unstable (via Hawking decay). The higher value has positive 
specific heat  $\partial M/\partial T> 0$, but if $T_0<T<T_1= 1/(\pi R)$, 
the black hole still has positive free energy. If $T>T_1$, the black 
hole is completely stable against decay.

For $AdS_5$, $T_0= \sqrt{2}/(\pi R)$, $T_1= 3/(2\pi R)$, 
$r_+=R/\sqrt{2}$, and at large 
M, $M\sim r_+^4/(w_4 R^2)$, thus 
\be
T\simeq \frac{(w_4M)^{1/4}}{\pi R^{3/2}}\sim M^{1/4}
\ee
whereas at small M (large R) we have approximately flat 5d space, 
\be
T\simeq \frac{1}{2\pi \sqrt{w_4M}}\sim M^{-1/2}
\ee
(for $AdS_4$ we would have $T\sim 1/(8\pi M)$). Here $w_4=8G_{N,5}/(3\pi)$.

Thus for $AdS_5$, we have T(M) going down from the maximal (initial)
temperature of $M_P$, initially with $T\sim M^{-1/2}$ as in flat 5d 
space, down to $T_0= \sqrt{2}/(\pi R)$, (an unstable branch), 
reached for $M_0=3R^2/4w_4$, then up
and completely stable above $T_1= 3/(2\pi R)$, and at infinity we have 
$T\sim M^{1/4}$.

Witten \cite{witten} has analyzed the thermodynamics of the system
of AdS plus black holes, by 
summing over contributing states, the dominating one being the one with 
lowest free energy.  AdS space dominates below $T_1$, and black holes 
dominate above $T_1$. Moreover, the lower M branch has larger free energy 
than the higher M branch, so if we study {\em equilibrium thermodynamics}, 
as Witten does, then we concentrate on the higher M branch. Moreover, in 
order to get on the AdS boundary a finite T field theory on $R^{n-1}\times S^1
$, Witten makes a scaling of the solution that involves high M, in 
which only the higher M black holes (and not the lower M, nor AdS) are 
relevant. 

But in the high energy collision process, which is a very non-equilibrium 
process, we expect to create the unstable black holes, as we do in flat 
space, not the stable, high mass black holes. Thus we expect {\em the lower 
mass branch of $T(M)$} (unstable one)
to be relevant, not the higher mass (stable one)
taken by Witten in the equilibrium case!

Otherwise however, if the black hole is sufficiently small, it should be 
created in  AdS space, and should not feel the IR cut-off, thus it should 
be the same AdS black hole. In the Appendix we have tried to make a 
coordinate transformation to a black hole that would be situated in Poincare
coordinates at $x_1$ (time) arbitrary, $x_0=b, \vec{x}=0$, with a horizon 
at $x_1$ arbitrary and $\vec{x}, x_0$ on a curve around the singularity, as we 
expect for the gravity dual scattering, on general grounds. We have not 
found a complete transformation, just sketched how one should do it, but 
shown that it is possible in principle.

Then the Witten analysis of the $T(M)$ curve should hold for $AdS_5$ 
even when we create black holes at fixed $\vec{x},x_0$. However, now we 
have a cut-off $AdS_5$, thus when the black hole becomes large enough it will 
reach the cut-off (IR brane). Eventually, when the black hole is large 
enough, it can be understood as black hole on the IR brane, and it will 
have a constant asymptotic temperature as a function of mass, 
\be
T=T_m=\frac{4aM_1}{\pi}
\label{irtemp}
\ee
 where a is yet not calculated, but should be of order one,
and $M_1= j_{1,1}/R\simeq 3.83/R$.

It is clear that the point when the black hole reaches the IR brane depends 
on the initial position $x_0$ of its center, which in realistic gravity 
duals depends on the details of the IR (but it is close to the IR) 
\cite{nastase2}. So as $\sqrt{s}$ is increased for the collision in the 
gravity dual, $T(M)$ for the produced black holes starts off in the lower 
branch of the AdS diagram, and then has a transition phase that depends 
on $x_0$, thus on the gravity dual details, after which it becomes 
asymptotically the $T(M)=T_1$=constant line, as in fig.\ref{temp}b.

But in \cite{kn,nastase2} it was shown that there are 3 main energy regions for
the scattering in the gravity dual, for $\sqrt{s}$  between $M_P=N^{1/4}
R^{-1}$  and $E_R=N^2R^{-1}$ the black holes are approximately in flat 
space, between $E_R(M_R)$ and a scale $E_F(M_F)$ 
depending on the gravity dual details 
in the IR, the black holes feel AdS space, and beyond $E_F(M_F)$ the black 
holes are efectively on the IR brane. But the lower branch for $T(M)$ in AdS
space terminates at $M=M_0$, and by then there are important AdS deviations 
from flat space, so we would like to have at least $M_0>E_R$. But 
$M_0= (9\pi/32)M_{P,5}^3R^2$, so it is time to remember that the cut-off
$AdS_5$  space was just a model, in reality we have also a compact space 
$X_5$. Then we would have $M_{P,5}^3= M_P^8{R'}^5$, where $R'$ is the size 
of $X_5$. In \cite{nastase2} it was found that for consistency of the dual 
gauge theory, in the IR one needs $R'\gg R$, which means that 
\be
M_0\sim M_{P,5}^3 R^2\sim M_P^8 {R'}^5R^2\gg M_P^8R^7=N^2R^{-1}=E_R
\ee
as needed. However, then our approximation of just cut-off $AdS_5$ as gravity
dual is inappropriate, and one would get a modified $T_0,M_0$ denoted 
$\tilde{T}_0,\tilde{M}_0$, and we will have the situation depicted in 
fig.\ref{temp}b.

\begin{figure}

\begin{center}

\includegraphics{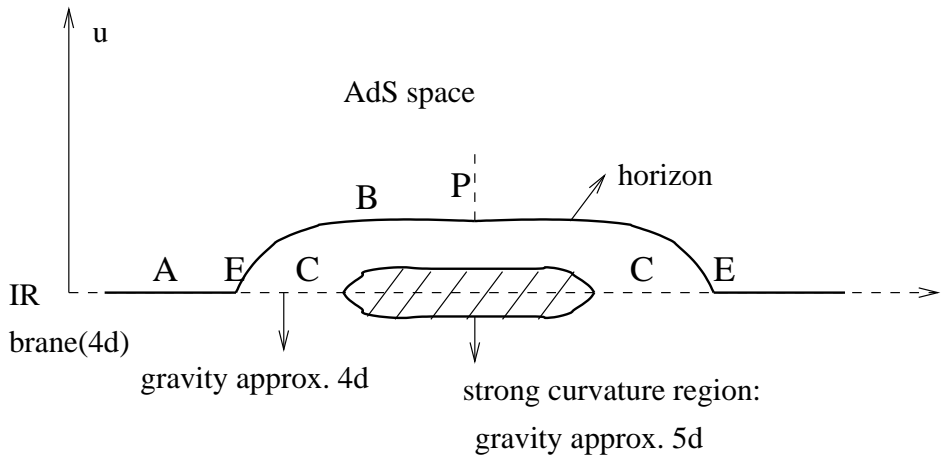}
\end{center}
\caption{Black hole on the IR brane. Region A is the asymptotic 4d IR brane,
region B is the asymptotic region for the horizon in AdS space (around
middle point P), same as for a black brane solution. The 4d edges E radiate
thermally. Inside the horizon, there is a low curvature region (gravity 
is approximately 4d), followed by a strong curvature region (shaded), 
where gravity is 5d, and should correspond to a new phase in gauge theory.}
\label{IRbrane}
\end{figure}

The solution for the black hole on the IR brane in the large mass limit 
was defined, outside the horizon, in \cite{amw}, and it looks like 
a pancake on the IR brane. It is depicted in fig.\ref{IRbrane}
 where we have also 
sketched what should happen inside the horizon (we will analyze that later). 
 The solution outside the 
horizon is interpolating between the gravity dual (AdS space with IR brane)
and the black brane solution (with translational invariance along the 
spatial IR brane coordinates) along the horizon, 
in the middle of the black hole (point P). The domain wall solution, in 
region A and B (along the horizon), considering that P goes over to infinity,
was calculated numerically in \cite{amw}. Because the black brane solution 
(that is a good approximation along the horizon around point P)
is translationally invariant and can be thought of as having constant 
energy density, \cite{amw} argued that $r_H\sim M^{1/3}$ (mass $\propto$
volume) in the asymptotic limit. But that is not necessarily so. 

The notion
of $r_H$ as well as the notion of mass refer to coordinates and metric valid 
asymptotically far away from the black hole, along the IR brane, not inside 
the black hole, along its horizon. In principle the two can be different.
A simple example of such a possibility is a collapsing star forced to 
collapse to a fixed density $\rho_0$, 
say $=M_P^4$: from the outside point of view,
the star never collapses inside the Schwarzschild radius, but slows down
exponentially,  thus its size remains $r_H=2GM\propto M$. But from the  
collapsing observer (at the star surface)
point of view, the star continues to collapse until it reaches 
$\rho=\rho_0$, thus a size of $r_0=(3M/4\pi \rho_0)^{1/3}\propto M^{1/3}$.
We will argue later that from the point of view of the faraway observer 
on the IR brane, the black hole has in fact $r_H\sim M^{1/2}$.

\section{ Thermodynamics: temperature and entropy of fireball}

The fireball is radiating mostly relativistic pions at fixed temperature. 
But for the pions radiated away, the entropy is proportional with the number 
of emitted pions. Indeed, 
for radiation at equilibrium composed of 
a relativistic particle species ($T\gg m, T\gg \mu$), 
\be
p=\frac{\rho}{3} \;\;\; \rho= \frac{\pi^2}{30} g T^4  \;\;\;
n= \frac{\zeta (3)}{\pi^2} g T^3 (boson)\;\;\;(
\rho= \frac{7}{8} \frac{\pi^2}{30} g T^4  \;\;\;
n=\frac{3}{4}\frac{\zeta (3)}{\pi^2} g T^3 (fermion))
\ee
and since the entropy of radiation is proportional to the volume, the entropy 
density is, for a boson (for a fermion, multiply by 7/8) 
\be
s=\frac{S}{V}= \frac{\rho +p}{T}=\frac{2\pi^2}{45} gT^3= \frac{2\pi^4}{45
\zeta (3)}n\simeq 3.60155 n
\ee

The emitted pions are not in equilibrium with the environment, but if the 
fireball is large enough, has fixed temperature and the pion emission doesn't
significally perturb the system, 
we can consider the pions to be in  equilibrium with 
the fireball (the radiating surface). If the emitted pions are relativistic,
we get for the total entropy of the pions emitted at constant temperature
 $S_{\pi}\simeq 3.6 N_{\pi}$. This will be corrected for finite fireballs
(for nonconstant temperature and perturbation of the system by emission).

The entropy of the emitted pions $S_{\pi}$ 
has to be larger than, but of the order of 
the initial entropy of the fireball, $S_{fireball}$. 
Therefore, if the fireball is dual to 
a black hole on the 4d IR brane, we expect the fireball entropy to also 
be equal to its area/4, in $\hat{M}_P$ units, with $\hat{M}_P$ the 
QCD analog of $M_P$, equal to $N^{1/3}\Lambda_{QCD}$, argued in \cite{nastase2}
to be experimentally about $1.3GeV$.  
Introducing a factor c for the difference between $S_{\pi}$ and $S_{fireball}$
 we should have asymptotically that 
\be
(S_{\pi}=) 
3.6 N_{\pi}= c\frac{\hat{M}_P^2 Area}{4}\simeq c\frac{(1.3 GeV)^2 Area}{4}
\label{arealaw}
\ee

This formula would then be a clear sign of black hole behaviour of the 
fireball.
But the area of the formed fireball is very hard (impossible?)
to measure experimentally, and even to define 
properly. One would need to measure the size of the fireball at a 
moment when it corresponds to a formed black hole.

But one can check a formula for the entropy that does not involve
the area, yet still contains black hole information.
 Indeed, for a Schwarzschild black hole, 
the entropy of emitted radiation time dt, occupying volume $dV=A c\, dt$ around
the black hole, is
\be
dS_{rad}= \frac{dU_{rad}+p_{rad}dV}{T}=\frac{4}{3}\frac{dU_{rad}}{T}
=\frac{4}{3}\frac{dM_{bh}}{T}=\frac{4}{3} dS_{bh}
\ee
since the emitted energy $dU_{rad}$ equals the energy lost by the black 
hole $dM_{bh}$. Since $T=1/(8\pi M)$, we get
\be
dS_{rad}=\frac{32 \pi }{3} M dM\Rightarrow S_{rad}= \frac{16\pi M^2}{3}= 
\frac{Area}{3}\;\;\; (Area= 4\pi (2M)^2)
\label{entropy}
\ee
thus the area law is recovered for the emitted radiation, 
but the coefficient is higher, which is 
still consistent with the second law, as the emitted entropy can be larger 
than the initial black hole entropy (Area/4), due to increase of total 
entropy during the emission process.

In other words, to check experimentally the area law for a small black hole
in flat 4d space, one could measure the total entropy of the emitted 
radiation and measure it against the intial area of the horizon similarly 
to (\ref{arealaw}). 
In QCD, the area law (\ref{arealaw}) would be a radical measurement of 
black hole like behaviour, since it involves a mass scale ($\hat{M}_P$), 
analog of $M_P$ for black holes, which for them 
signals quantum gravity behaviour.
In \cite{nastase} another test of dual black hole physics
was noted, for the saturation of the 
conjectured bound for viscosity over entropy density, $\eta/s\geq 1/(4\pi)$
\cite{kss},
that is experimentally very close to the theoretical value \cite{shuryak},
but that did not involve $M_P$, so it was maybe easier to understand why 
a gauge theory object would saturate it as well as a black hole. Here 
on the other hand, $M_P$ appears explicitly in the area law
on the dual side, and becomes 
$\hat{M}_P=N^{1/4}\Lambda_{QCD}$ in gauge theory. 

A less ambitious measurement for the entropy of a small black hole in 
flat space would be (according to (\ref{entropy})) to match the 
entropy of the emitted radiation against $M^2$ ($M$= initial mass of black 
hole). Of course, that would measure the area law only if we knew that 
the horizon size scales with $M$, and as it is it seems that we only measure 
$\int dM/T(M)$, but this in itself assumes certain properties about the 
black hole. Specifically, that it is a truly blackbody thermal object within 
a good accuracy, and the radiated particles are relativistic. Under these 
assumptions, measuring $S_{rad}(M)$ is the same as measuring $T(M)$ over a
large range of masses.

Of course, experimentally, even measuring $T(M)$ over a sufficiently 
range of masses is 
harder for a small black hole, so the $S(M)$ measurement is probably better,  
as well as proving the above properties.

Thus for the case of the black hole on the IR brane, when we have conjectured
\cite{nastase} that the temperature should be constant and equal to 
$T=4aM_1/\pi$, we should get 
\be
S_{rad}=\frac{4}{3T}M_{bh}= \frac{\pi}{3a}\frac{M_{bh}}{M_1}
\ee

Correspondingly, for the fireball we expect 
\be
S_{\pi}=3.6N_{\pi}=\frac{\pi}{3a}\frac{M_{fireball}}{m_{\pi}}
\ee
and at constant $\sqrt{s}/nucleon=200GeV/nucleon$, 
the fireball mass (proportional to the 
effective $\sqrt{s}$ of the reaction) is proportional to 
the number of nucleons participating in the reaction, ${\cal A}$.
So we expect that $N_{\pi}$ is proportional to ${\cal A}$.

Let us now turn to the temperature of the fireball. It was argued in 
\cite{nastase} that the temperature of the black hole on the IR brane
should be be (\ref{irtemp}). Since $dM=T(M)dS=T(M)d(Area)/4$, to have $T(M)$
constant, we need that the area of the horizon scales like $M$. Here 
horizon area refers to the 2d area visible from the 4d IR brane, i.e. the 
edges E of the black hole in fig.\ref{IRbrane}, 
that is the only part that Hawking
radiates (otherwise we would not have a proper 4d limit for gravity
and Hawking's 4d calculation would be meaningless if it could be embedded in 
higher dimensional theories). Thus $r_H(M)\sim M^{1/2}$ and constant 
temperature $T(M)$ are equivalent. More precisely, for (\ref{irtemp})
one needs
\be
M_P 2 r_H=\sqrt{\frac{M}{aM_1}}
\label{horizon}
\ee
We have seen in the introduction that not only the cross section for black 
hole formation, but also the area of the trapped surface formed at zero 
impact parameter go like $ln^2\sqrt{s}$, whereas the perturbative area 
of the horizon of a black hole on the IR brane goes like $ln^2 M_{bh}$.
However, the cross section for scattering is not necessarily related to the 
black hole horizon, whereas the area of the trapped surface is only bound 
to be smaller than the area of the horizon of the black hole that forms, 
and a power law is larger than $ln^2\sqrt{s}$. And perturbation theory 
has no reason to be valid close to the horizon. 

To understand this scaling, let's look at what happens inside the horizon 
of the large 
IR brane black hole (fig.\ref{IRbrane}).  As one crosses the horizon, nothing 
special should happen (same as for a large Schwarzschild black hole, 
the horizon has small curvature, so an infalling observer doesn't feel 
anything at the horizon). Thus one should still have a massive 4d 
gravitational interaction between masses, $\delta 
h_{00}\sim G_4\delta Me^{-M_1r}/r$.
As one falls into a Schwarzschild black hole, one eventually reaches the 
strong curvature region (the singularity). Similarly now, we expect that 
in the center of the black hole we have a strong curvature region (but no 
singularity, as argued before). The strong curvature region will exhibit 
the full 5d gravity (we cannot consider the dimensionally reduced version 
to 4d massive gravity anymore). Correspondingly, now perturbative gravity 
changes as 
\be
h_{00}\propto \frac{Me^{-M_1r}}{M_{P,4}^2r}\rightarrow 
\frac{M}{M_{P,5}^3r^2}=\frac{M}{M_{P,4}^2R^{-1}r^2}\propto \frac{M}{M_{P,4}M_1
r^2}
\ee
since $M_1=j_{1,1}R^{-1}\simeq 3.83 R^{-1}$. If we then consider that in 
the asymptotic (large M) limit the 
size of the black hole is determined by the size of the strong curvature, 5d 
gravity region, the perturbative result for the horizon ($h_{00}\sim 1$)
would give indeed 
\be
M_{P,4}^2 4 r_H^2\propto \frac{M}{M_1}
\ee
as advocated (with $M_{P,4}$ becoming $\hat{M}_{P}=N^{1/4}\Lambda_{QCD}$ in 
QCD). In flat 5 dimensions, the horizon size (where $h_{00}\sim 1$
for a point mass $M$) would give a constant of proportionality $a=32 j_{1,1}/
3\pi$ instead of 1, but we expect just the scaling behaviour to still be 
valid anyway. 

What would this picture mean for the dual fireball? The center of the 
fireball should be dual to the center of the black hole on the IR brane, and 
small perturbations of fixed size should
 interact with a $1/r^2$ Yang-Mills potential, dual
to the 5d gravitational potential. Thus this state could correspond to a 
new state of matter, maybe the CGC (color glass condensate) state assumed in 
some QCD models to dominate at the early stages of the collision. 
Outside the core, close to the surface of the fireball, we have
the same Coulomb potential described in \cite{nastase}, for free quarks and 
gluons, dual to the 4d Newtonian potential of small perturbations inside 
region C of the black hole on the IR brane. 

In the realistic case of the pion mass being the lightest, the pion field
profile (\ref{dbiscalar}) dominates, and it also acts as a black hole, as 
we discussed. Its horizon is at (if the DBI scale $\Lambda\sim \hat{M}_P$)
$\hat{M}_P^2r_0^2\sim \Lambda^2 r_0^2=\bar{C}$. But $\bar{C}$ is a scalar 
charge (which can be seen from the large r asymptotics $\phi(r)\sim \bar{C}
/r$, see \cite{nastase3} for a more complete treatment), and if we have a 
fireball of mass $M$ decaying mostly into pions, the total (quantized) 
scalar charge is $\bar{C}\simeq M/m_{\pi}$, thus 
\be
\hat{M}_P^2 r_0^2\sim \frac{M}{m_{\pi}};\;\;\; \hat{M}
_P2r_0\equiv \sqrt{\frac{M}{am_{\pi}}} 
\ee
Then the same argument as for a black hole gives for the pion field 
distribution
\be
 dM = T dS=\frac{\pi T}{4}\hat{M}_P^24 dr_0^2\Rightarrow
 dM=\frac{\pi T}{4} \frac{dM}{am_{\pi}}\Rightarrow T=\frac{4a}{\pi}m_{\pi}
\ee
that is, the same formula put forward in \cite{nastase}, but now it is 
derived from a purely effective field theory (pion) model.  Of course, we 
used the black hole information that even for the pion field profile, the 
entropy is = area/4, in $\hat{M}_P$ units. For the pure Yang-Mills case, 
dual to a black hole, that was clear, but now we have to assume the same 
will continue to be true even if we introduce light pions (light quarks).
This is very likely, since for instance in the black hole case we could 
keep $M_P=N^{1/4}R^{-1}$ fixed, but take $M_1\sim R^{-1}$ (the mass of 
emitted particles) to zero and still have $S= M_P^2 Area/4$, so most 
likely the same will happen if brane bending (dual to light pions) dominates 
the dynamics and $m_{\pi}$ (mass of emitted particles) is small. 

In conclusion, we have seen that the information paradox is in fact the 
same for black holes and for the creation and decay of fireballs at RHIC, 
we have seen that apparent contradictions of the QCD and dual picture 
for the fireball can be resolved. The black holes produced in the gravity 
dual correspond to the unstable, lower mass branch of the AdS black holes 
considered by Witten for finite temperature gauge theory, and we found 
how $T(M)$ should look like. We have derived the area law for the entropy 
of the fireball. A less ambitious formula for the fireball entropy gives a 
linear relation for the number of emitted pions vs. the 
C.M. energy ($\sqrt{s}$) involved in the reaction. We gave more arguments for 
the fact that $T_{fireball}\propto m_{\pi}$, including an argument based on the
simple DBI scalar model for the pion. Black holes on the IR brane 
should have a 5d strong curvature region, that may correspond to a new state
of matter deep inside the fireball (or at its early stages).

{\bf Acknowledgements} I would like to thank 
Robert de Mello Koch for discussions and for a critical reading of the 
manuscript. I would also like to thank Ofer Aharony, Antal Jevicki,
 Dmitri Kharzeev, Robert Pisarski,
Edward Shuryak, Dam Thanh Son, Lenny Susskind,   Chung-I 
Tan for discussions.
This research was  supported in part by DOE
grant DE-FE0291ER40688-TaskA.

\newpage

{\Large\bf{Appendix A. AdS black hole solution }}

\renewcommand{\theequation}{A.\arabic{equation}}
\setcounter{equation}{0}

\vspace{1cm}

Here we will study the AdS black hole considered by Witten \cite{witten}
and how it looks in different coordinate systems. 
The metric of global $AdS_{n+1}$ space can be written as 
\be
ds^2= -(\frac{r^2}{R^2}+1)dt^2 +\frac{dr^2}{\frac{r^2}{R^2}+1}+ r^2 d\Omega^2
=-(1+u^2)dt^2+d\vec{u}d\vec{u}-\frac{u^2du^2}{1+u^2}
\label{ads}
\ee
where $R=\sqrt{-\Lambda/3}$.
With the transformation of coordinates $r/R=sinh \; \rho = tan \; \bar{
\rho}$ one obtains
\bea
ds^2&=&-dt^2 \; cosh^2 \rho +d\rho^2 +sinh^2\rho d\Omega^2\nonumber\\
&=& \frac{1}{\cos ^2 \bar{\rho}}(-dt^2 +d\bar{\rho}^2 +\sin^2 \bar{\rho} 
d\Omega^2)
\eea
From the original coordinates, the transformation
\be
e^{\frac{t}{b}}=\sqrt{x_0^2+\vec{x}^2}= x_0 \sqrt{1+\frac{r^2}{R^2}};\;\;\;
\frac{r}{R}\vec{\Omega}=\vec{u}=\frac{\vec{x}}{x_0};\;\;\; x_0=e^y
\label{coordtr}
\ee
gives the Poincare patch of AdS space
\be
ds^2= R^2 \frac{d\vec{x}^2+dx_0^2}{x_0^2}= R^2 [e^{-2y}d\vec{x}^2+dy^2]
\label{poincare}
\ee

The AdS black hole in global coordinates just replaces
\be
V=1+\frac{r^2}{R^2}\rightarrow 1+\frac{r^2}{R^2}-\frac{w_n M}{r^{n-2}}
\label{bh}
\ee
where $w_n=16\pi G_{N,5}/((n-1)\Omega_{n-1})$. So the question is, how does 
this black hole look like in Poincare coordinates, that define the ``cut-off
AdS'' by $y\rightarrow -|y|$. 

The problem is that there is a certain ambiguity in the coordinate 
transformation to Poincare-like coordinates. 
We will first try the original coordinate 
transformation (\ref{coordtr}) (for the AdS background), but it is not 
clear that this is relevant for the black hole case. We can modify the 
coordinate transformation such that it asymptotes to the usual AdS 
transformation only when the AdS black hole asymptotes to AdS. 

If we take apply (\ref{coordtr}) on the black hole (\ref{bh}), we get 
a horizon situated at
\be
r=r_+\Rightarrow x= \frac{r_+}{R} x_0= \frac{r_0}{R} e^y
\ee
and the physical region is at $r\geq r_+, x\geq r_+ x_0/R$. In Euclidean 
space, we cannot go beyond that, but in a Minkowski continuation, we 
would have a real singularity at $r=0$. In the Poincare patch coordinates, the 
singularity would be at $x=0, x_0=$arbitrary, i.e. the y axis. 
Note that for the absence of singularities, we needed a periodicity in t,
$t\equiv t+\beta_0$, translating now into $y\equiv y+\beta_0$ ($x_0\equiv
x_0 e^{\beta_0}$), and the identifications are made along lines parallel to 
the horizon.
But this is not what we want.

In A-S shockwave collisions, we expect to create a black hole situated at
(i.e.,  with singularity at) $y=y_0$ ($x_0=$fixed). Moreover, the IR brane 
will be situated by definition at $x_0=1$ ($y=0$). But the AdS black hole 
has a singularity that is instead at arbitrary $x_0$ and fixed $r$ (=0).
So we need to modify the 
coordinate transformation (\ref{coordtr}) accordingly. 

We want the singularity to be not at arbitrary $x_0$, but rather at 
arbitrary euclidean time $x_1$ (one of the $\vec{x}$ coordinates 
that will be rotated to Minkowski space).
The original coordinate transformation keeps $\Omega_d$ intact, but as we
want now to get $(r,t)(x, x_1, x_0)$, as opposed to $(r,t)(x, x_0)$ for 
the background, we need to break $\vec{\Omega}_d$ as $(\cos \theta, \sin\theta
\vec{\Omega}_{d-1})$. Then we would need to provide also a $
\theta(x, x_1, x_0)$ to complete the transformation, and it probably should 
be checked that such a transformation exists. 

An example of a transformation that gives a singularity at $x_0=R$ (1 in 
rescaled notation), $\vec{x}'=0, x_1=$arbitrary and reduces at $r=\infty$
to the usual transformation is
\bea
 \frac{r^2}{R^2}&=&\frac{\vec{x}^2-x_1^2}{x_0^2}(1+\frac{w_n M}{r^{n-2}})
+ \frac{x_1^2}{x_0^2}\frac{1}{1+\frac{w_n M}{r^{n-2}}}\nonumber\\
e^{2t}&=&x_0^2+\vec{x}^2+(x_0^2 +\vec{x}^2 -x_1^2 -R^2)\frac{w_n M}{r^{n-2}}
\nonumber\\&=&R^2+x_1^2 +(x_0^2-R^2)(1+\frac{w_n M}{r^{n-2}})+(\vec{x}^2-
x_1^2)(1+\frac{w_n M}{r^{n-2}})
\eea

Unfortunately, for this coordinate transformation the horizon is at
\bea
&&A(\vec{x}^2-x_1^2)+\frac{x_1^2}{A}= (A-2)x_0^2;\;\;\; A\equiv 
2+\frac{r_+^2}{R^2}\nonumber\\&&
A(\vec{x}^2-x_1^2 +x_0^2 -R^2)+x_1^2 +R^2 = e^{2t}=arb.
\eea
thus if $x_1$ is arbitrary, $x_0$ is arbitrary as well, and in any case
$(\vec{x}^2-x_1^2)$ increases with $x_0^2$. 

That can be cured by taking the following transformation
\bea
&&\frac{r^2}{R^2}(1+\frac{d}{r}\frac{x_1^2+R^2}{x_0^2})=\frac{\vec{x}^2-x_1^2}
{x_0^2} (1+\frac{w_nM}{r^{n-2}})+\frac{x_1^2/x_0^2 + c x_0^2/r^{1/2}}{
1+\frac{w_nM}{r^{n-2}}}\nonumber\\&&
e^{2t}=x_0^2+\vec{x}^2+(x_0^2 +\vec{x}^2 -x_1^2 -R^2)\frac{w_n M}{r^{n-2}}
\eea
Then by requiring that the horizon is an ellypse
in $\vec{x}'$ and $x_0^2$ around $\vec{x}'=0, x_0^2=R^2$,
we get 
\be
c= \frac{A(A-2)r_+^{1/2}}{2R^2};\;\;\; d= \frac{r_+}{A(A-2)}
\ee
With this transformation, we have the usual AdS transformation both at 
$r=\infty$ and at $M=0$, the singularity is at $\vec{x}'=0, x_0=R, 
x_1=$arbitrary and the horizon at $x_1$ arbitrary and $\vec{x}'$ and $x_0$
on the ellypse
\be
A(\vec{x}^2-x_1^2)+\frac{A-2}{2R^2}(x_0^2-R^2)^2 =\frac{R^2}{A}+\frac{A-2}
{2}R^2
\ee

\end{document}